\pdfoutput=1

\documentclass[11pt]{article}

\usepackage{emnlp2021}

\usepackage{times}
\usepackage{latexsym}
\usepackage{amssymb}
\usepackage{amsmath}
\usepackage{graphicx}
\usepackage{subfigure}
\usepackage{multirow}

\usepackage[T1]{fontenc}

\usepackage[utf8]{inputenc}

\usepackage{microtype}

\title{Efficient Document Retrieval by End-to-End Refining and Quantizing BERT Embedding with Contrastive Product Quantization}

\author{}
\author{Zexuan Qiu$^{1}$, Qinliang Su$^{1, 2}$\thanks{$^{*}$Corresponding author.}, Jianxing Yu$^{3}$, and Shijing Si$^{4}$\\
 $^1$School of Computer Science and Engineering, Sun Yat-sen University, Guangzhou, China \\
 $^2$ Guangdong Key Laboratory of Big Data Analysis and Processing, Guangzhou, China\\
 $^3$School of Artificial Intelligence, Sun Yat-sen University, Guangzhou, China\\
 $^4$School of Economics and Finance, Shanghai International Studies University, China\\
 {\tt\{qiuzx3@mail2, suqliang@mail, yujx26@mail\}.sysu.edu.cn}
 }

\begin{document}
\maketitle
\begin{abstract}

Efficient document retrieval heavily relies on the technique of semantic hashing, which learns a binary code for every document and employs Hamming distance to evaluate document distances. However, existing semantic hashing methods are mostly established on outdated TFIDF features, which obviously do not contain lots of important semantic information about documents. Furthermore, the Hamming distance can only be equal to one of several integer values, significantly limiting its representational ability for document distances. To address these issues, in this paper, we propose to leverage BERT embeddings to perform efficient retrieval based on the product quantization technique, which will assign for every document a real-valued codeword from the codebook, instead of a binary code as in semantic hashing. Specifically, we first transform the original BERT embeddings via a learnable mapping and feed the transformed embedding into a probabilistic product quantization module to output the assigned codeword. The refining and quantizing modules can be optimized in an end-to-end manner by minimizing the probabilistic contrastive loss. A mutual information maximization based method is further proposed to improve the representativeness of codewords, so that documents can be quantized more accurately. Extensive experiments conducted on three benchmarks demonstrate that our proposed method significantly outperforms current state-of-the-art baselines\footnote{Our PyTorch code is available at \url{https://github.com/qiuzx2/MICPQ}, and our MindSpore code will be also released soon.}.

\end{abstract}

\section{Introduction}
In the era of big data, Approximate Nearest Neighbor (ANN) search has attracted tremendous attention thanks to its high search efficiency and extraordinary performance in modern information retrieval systems. By quantizing each document as a  compact binary code, semantic hashing \citep{SalakhutdinovH09} has become the main solution to ANN search due to the extremely low cost of calculating Hamming distance between binary codes. One of the main approaches for unsupervised semantic hashing methods is established on generative models \citep{ChaidaroonF17,HenaoCSSWWC18,DongSSC19,ZhengSSC20}, which encourage the binary codes to be able to reconstruct the input documents. Alternatively, some other methods are driven by graphs \citep{WeissTF08, chaidaroon2020node2hash,Hansen0SAL20a, OuSYLWZCZ20}, hoping the binary codes can recover the neighborhood relationship. Though these methods have obtained great retrieval performance, there still exist two main problems.

Firstly, these methods are mostly established on top of the outdated TFIDF features, which do not contain various kinds of important information of documents, like word order, contextual information, etc. In recent years, pre-trained language models like BERT have achieved tremendous success in various downstream tasks. Thus, a natural question to ask is whether we can establish efficient retrieval methods on BERT embeddings. However, it has been widely reported that BERT embeddings are not suitable for semantic similarity-related tasks \citep{ReimersG19}, which perform even worse than the traditional Glove embeddings \citep{PenningtonSM14}. \citep{ethayarajh2019contextual, LiZHWYL20} attribute 
this to the "anisotropy" phenomenon that BERT embeddings only occupy a narrow cone in the vector space, causing the semantic information hidden in BERT embeddings not easy to be leveraged directly. Thus, it is important to investigate how to effectively leverage the BERT embeddings for efficient document retrieval.

Secondly, to guarantee the efficiency of retrieval, most existing methods quantize every document to a binary code via semantic hashing. There is no doubt that the Hamming distance can improve the retrieval efficiency significantly, but it also restricts the representation of document similarities seriously since it can only be an integer from $-B$ to $B$ for $B$-bit codes. Recently, an alternative approach named product quantization \citep{JegouDS11,JangC21,abs-2109-05205} has been proposed in the computer vision community to alleviate this problem. Basically, it seeks to quantize every item to one of the codewords in a codebook, which is represented by a Cartesian product of multiple small codebooks. It has been shown that product quantization is able to deliver superior performance than semantic hashing while keeping the cost of computation and storage relatively unchanged. However, this technique has rarely been explored in unsupervised document retrieval.

Motivated by the two problems above, in this paper, we propose an end-to-end contrastive product quantization model to jointly refine the original BERT embeddings and quantize the refined embeddings into codewords. Specifically, we first transform the original BERT embeddings via a learnable mapping and feed the transformed embedding into a probabilistic product quantization module to output a quantized representation (codeword). To preserve as much semantic information as possible in the quantized representations, inspired by recent successes of contrastive learning, a probabilistic contrastive loss is designed and trained in an end-to-end manner, simultaneously achieving the optimization of refining and quantizing modules. Later, to further improve the retrieval performance, inspired by the recent development of clustering, a mutual information maximization based method is further developed to increase the representativeness of learned codewords. By doing so, the cluster structure hidden in the dataset of documents could be kept soundly, making the documents be quantized more accurately. Extensive experiments are conducted on three real-world datasets, and the experimental results demonstrate that our proposed method significantly outperforms current state-of-the-art baselines. Empirical analyses also demonstrate the effectiveness of every proposed component in our proposed model.

\section{Preliminaries of Product Quantization for Information Retrieval}
\label{Sec:Preliminary}

In fields of efficient information retrieval, a prevalent approach is semantic hashing, which maps every item $x$ to a binary code $b$ and then uses Hamming distances to reflect the semantic similarity of items. Thanks to the low cost of computing Hamming distance, the retrieval can be performed very efficiently. However, the Hamming distance can only be an integer from $-B$ to $B$ for $B$-bit codes, which is too restrictive to reflect the rich similarity information.

An alternative approach is vector quantization (VQ) \citep{gray1998quantization}, which assigns every item with a codeword from a codebook $C$. The codeword in VQ could be any vector in ${\mathbb{R}}^D$, rather than limited to the binary form as in semantic hashing. By storing pre-computed distances between any two codewords in a table, the distance between items can be obtained efficiently by looking up the table. However, to ensure competitive performance, the number of codewords in a codebook needs to be very large. For example, there are $2^{64}$ different codes for a 64-bit binary code, and thus the number of codewords in VQ should also be of this scale, which however is too large to be handled.

To tackle this issue, product quantization \citep{JegouDS11} proposes to represent the codebook $C$ as a Cartesian product of $M$ small codebooks
\begin{equation}
    C = C^1 \times C^2 \times \cdots \times C^M,
\end{equation}
where the $m$-th codebook $C^m$ consists of $K$ codewords $\{c^m_k\}_{k=1}^K$ with $c^m_k \in {\mathbb{R}}^{D/M}$. For an item, the product quantization will choose a codeword from every codebook $C^m$, and the final codeword assigned to this item is
\begin{equation}
    c = c^1 \ \circ \ c^2  \cdots \circ \ c^M,
\end{equation}
where $c^m$ denotes the codeword chosen from $C^m$, and $\circ$ denotes concatenation. For each codeword $c$, we only need to record its indices in the $M$ codebooks, which only requires $M\log_2 K$ bits. Thanks to the Cartesian product decomposition, now we only need to store $MK$ codewords of dimension ${\mathbb{R}}^{D/M}$ and $M$ lookup tables of size $K\times K$. As an example, to enable a total of $2^{64}$ codewords, we can set $M=32$ and $K=4$, which obviously will reduce the size of footprint and lookup tables significantly. During retrieval, we only need to look up the $M$ tables and sum them up, which is only slightly more costly than the computation of Hamming distance.

\section{The End-to-End Joint Refining and Quantizing Framework}

To retrieve semantic-similar documents, a core problem is how to produce for every document a quantized representation that preserves the semantic-similarity information of original documents. In this section, a simple two-stage method is first proposed, and then we develop an end-to-end method that is able to directly output representations with desired properties.

\subsection{A Simple Two-Stage Approach}
To obtain semantics-preserving quantized representations, a naive approach is to first promote the semantic information in original BERT embeddings and then quantize them. Many methods have been proposed on how to refine BERT embeddings to promote their semantic information. These methods can be essentially described as transforming the original embedding $z(x)$ into another one $\tilde z(x)$ 
via a mapping $g(\cdot)$ as
\begin{equation}
    \tilde z(x) = g(z(x)),
\end{equation}
where $g(\cdot)$ could be a flow-based mapping \cite{LiZHWYL20}, or a neural network trained to maximize the agreement between representations of a document's two views \cite{GaoYC21}, etc. It has been reported that the refined embeddings $\tilde z(x)$ are semantically much richer than original one $z(x)$. Then, we can follow standard product quantization procedures to quantize the refined embeddings $\tilde z(x)$ into discrete representations.

\subsection{End-to-End Refining and Quantizing via Contrastive Product Quantization}

Obviously, the separation between the refining and quantizing steps in the two-stage method could result in a significant loss of semantic information in the final quantized representation. To address this issue, an end-to-end refining and quantizing method is proposed. We first slice the original BERT embedding $z(x)$ into $M$ segments $z^m(x) \in {\mathbb{R}}^{D/M}$ for $m=1, 2, \cdots, M$. Then, we refine $z^m(x)$ by transforming it into a semantic-richer form $\tilde z^m(x)$ via a mapping $g^m_\theta(\cdot)$, that is,
\begin{equation}
    \tilde z^m(x) = g^m_\theta(z^m(x)),
\end{equation}
where the subscript $\theta$ denotes the learnable parameter. Different from the mapping $g(\cdot)$ which is determined at the refining stage and is irrelevant to the quantization in the two-stage method, the mapping $g^m_\theta(\cdot)$ here will be learned later by taking the influences of quantization error into account. Now, instead of quantizing the refined embedding $\tilde z^m(x)$ to a fixed codeword,  we propose to 
 quantize it to one of the codewords $\{c_{k^m}^m\}_{k^m=1}^K$ by stochastically selecting $k^m$ according to the  distribution
\begin{equation} \label{pk_given_m}
    p(k^m|x) = \frac{\exp\left(-\left\| \tilde z^m(x) - c^m_{k^m}\right\|^2\right)}{\sum_{i=1}^K\exp\left(-\left\| \tilde z^m(x) - c^m_i\right\|^2\right)}
\end{equation}
with $k^m=1, 2, \cdots, K$. Obviously, the probability that $\tilde z^m(x)$ is quantized to a codeword is inversely proportional to their distance. Thus, by denoting $k^m$ as a random sample drawn from $p(k^m|x)$, {\it i.e.}, $k^m \sim p(k^m|x)$, we can represent the $m$-th quantized representation of document $x$ as 
\begin{equation} \label{h_codeword_sel}
    h^m(x) = C^m \cdot one\_hot(k^m),
\end{equation}
and the whole quantized representation of $x$ as
\begin{equation} \label{h_concat}
   h(x) = h^1(x) \circ h^2(x) \cdots \circ h^M(x).
\end{equation}
Note that the quantized representation $h(x)$ depends on random variables $k^m \sim p(k^m|x)$ for $m=1, 2, \cdots, M$, thus $h(x)$ itself is also random. 

Now, we seek to preserve as much semantic information as possible in the quantized representation $h(x)$. Inspired by the recent successes of contrastive learning in semantic-rich representation learning \cite{GaoYC21}, we propose to minimize the contrastive loss. Specifically, for every document $x$, we first obtain two BERT embeddings by passing it through BERT two times with two independent dropout masks and then use the embeddings to generate two quantized representations $h^{(1)}(x)$ and $h^{(2)}(x)$ according to \eqref{h_codeword_sel} and \eqref{h_concat}. Then, we define the contrastive loss as
\begin{equation}
    {\mathcal{L}}_{cl} = - \frac{1}{|{\mathcal{B}}|} \sum_{x \in {\mathcal{B}}} \left( \ell^{(1)}(x) + \ell^{(2)}(x) \right),
\end{equation}
where ${\mathcal{B}}$ denotes a mini-batch of training documents; and $\ell^{(i)}(x)$ for $i=1, 2$ is defined as
\begin{equation}
    \ell^{(i)}(x) \triangleq \log \frac{{\mathcal{S}}(h^{(1)}_x, h^{(2)}_x)}{{\mathcal{S}}(h^{(1)}_x, h^{(2)}_x)  + \!\!\!\! \sum\limits_{\substack{t\in {\mathcal{B}} \backslash x \\ n = 1,2}} \!\!\! {\mathcal{S}}(h^{(i)}_x, h^{(n)}_t)},
\end{equation}
with $h^{(1)}_x$ denoting the abbreviation of $h^{(1)}(x)$ for conciseness; and ${\mathcal{S}}(h_1, h_2)$ is defined as
\begin{equation}
    {\mathcal{S}}(h_1, h_2) \triangleq \exp\left(sim(h_1, h_2)/\tau_{cl}\right),
\end{equation}
with $sim(h_1, h_2) \triangleq \frac{h_1^T h_2}{\left\|h_1\right\| \left\|h_2\right\|}$ being the cosine similarity function, and $\tau_{cl}$ being a temperature hyper-paramter.

Under the proposed quantization method above, the quantized representation $h(x)$ depends on the random variable $k^m \sim p(k^m|x)$, making it not deterministic w.r.t. a given document $x$. Thus, we do not directly optimize the random contrastive loss ${\mathcal{L}}_{cl}$, but minimize its expectation
\begin{equation} \label{L_expectation}
    \overline {\mathcal{L}}_{cl} = - \frac{1}{|{\mathcal{B}}|} \sum_{x \in {\mathcal{B}}} \left( \overline \ell^{(1)}(x) + \overline \ell^{(2)}(x) \right),
\end{equation}
where $\overline \ell^{(i)}(x)$ represents the expectation of $\ell^{(i)}(x)$ w.r.t. $k^m \sim p(k^m|x)$ for $m=1, 2, \cdots, M$, that is,
\begin{equation} \label{ell_expectation}
    \overline \ell^{(i)}(x) = {\mathbb{E}}_{k^1, k^2, \cdots, k^M} \left[ \ell^{(i)}(x) \right].
\end{equation}
Obviously, it is impossible to derive an analytical expression for $\overline \ell^{(i)}(x)$, making the optimization of $\overline {\mathcal{L}}_{cl}$ not feasible. To address this issue, it has been proposed in \citep{JangGP17} that the random sample $k^m$ drawn from distribution $p(k|x) = \frac{\exp\left(-\left\| \tilde z^m(x) - c^m_{k}\right\|^2\right)}{\sum_{i=1}^K\exp\left(-\left\| \tilde z^m(x) - c^m_i\right\|^2\right)}$ can be re-parameterized as
\begin{equation}
    k^m = \arg \max_{i} \left[ - \left\| \tilde z^m(x) - c^m_{i}\right\|^2 + \xi_i \right],
\end{equation}
where $\xi_i$ denote i.i.d. random samples drawn from the Gumbel distribution $Gumbel(0, 1)$. Then, using the softmax function to approximate the $\arg \max(\cdot)$, the $m$-th quantized representation $h^m(x)$ can be approximately represented as
\begin{equation}
    \widetilde h^m(x) = C^m \cdot v,
\end{equation}
where $v \in {\mathbb{R}}^K$ is a probability vector whose $k$-th element is
\begin{equation} \label{v_k_expression}
    v_k = \frac{\exp\big(- \frac{\left\| \tilde z^m(x) - c^m_{k}\right\|^2 + \xi_k }{\tau} \big)}{\sum_{i=1}^K \exp\big(- \frac{\left\| \tilde z^m(x) - c^m_{i}\right\|^2 + \xi_i }{\tau}\big)},
\end{equation}
with $\tau$ being a hyper-parameter. It can be easily seen that $\widetilde h^m(x)$ will converge to $h^m(x)$ as $\tau \rightarrow 0$, thus $\tilde h^m(x)$ can be used as a good approximation to $h^m(x)$. With the approximation, $\overline \ell^{(i)}(x)$ in \eqref{ell_expectation} can be approximately written as
\begin{equation} \label{ell_expectation_approx}
    \overline \ell^{(i)}(x) \approx \log \frac{{\mathcal{S}}(\widetilde h^{(1)}_x, \widetilde h^{(2)}_x)}{{\mathcal{S}}(\widetilde h^{(1)}_x, \widetilde h^{(2)}_x)  + \!\!\!\! \sum\limits_{\substack{t\in {\mathcal{B}} \backslash x \\ n = 1,2}} \!\!\! {\mathcal{S}}(\widetilde h^{(i)}_x, \widetilde h^{(n)}_t)},
\end{equation}
where $\widetilde h_x^{(1)}$ is the abbreviation of $h^{(1)}(x)$. Substituting \eqref{ell_expectation_approx} into \eqref{L_expectation} gives an approximate analytical expression of $\overline L_{cl}$ that is differentiable w.r.t. the parameter $\theta$ for refining and codebooks $\{C^m\}_{m=1}^M$ for quantization. Therefore, we can optimize the $\theta$ and codebooks $\{C^m\}_{m=1}^M$ in an end-to-end manner, explicitly encouraging the quantized representations $h(x)$ to preserve more semantic information.

It is worth noting that the injected Gumbel noise $\xi_i$ in \eqref{v_k_expression} is important to yield a sound approximation  $\overline \ell^{(i)}(x)$ in \eqref{ell_expectation_approx}. Theoretically, the approximated $\overline \ell^{(i)}(x)$ is guaranteed to converge to the exact value when $\tau \to 0$ and a large number of independent Gumbel nosies are used to approximate the expectation. However, if we abandon this noise in the computation of $v_k$, we will lose the appealing property above. Our experimental results also demonstrate the advantages of injecting Gumbel noises in the approximation. Another point worth pointing out is that if the refined embedding $\tilde z^m(x)$ is quantized to the closest codeword deterministically, that is, letting $k^m = \arg \max_{i} \left[ - \left\| \tilde z^m(x) - c^m_{i}\right\|^2 \right]$, then it becomes equivalent to our probabilistic quantization approach without using Gumbel noise, further demonstrating the advantages of our method.

\subsection{Improving the Representativeness of Codewords via MI Maximization}

It can be seen that the codewords in $C^m$ work similarly to the cluster centers in clustering. The clustering centers are known to be prone to get stuck at suboptimal points, which applies to the codewords analogously. If the codewords are not representative enough, they could result in a significant loss of semantic information in the quantized representations \citep{GeHK013,CaoL0ZW16}.

It has been recently observed that maximizing mutual information (MI) between the data and the cluster assigned to it can often lead to much better clustering performance \citep{HuMTMS17,JiVH19,Do0V21}. Inspired by this, to increase the representativeness of codewords, we also propose to maximize the MI between the original document $x$ and the codeword (index) assigned to it. To this end, given the conditional distribution $p(k^m|x)$ in \eqref{pk_given_m}, we first estimate the marginal distribution of codeword index $k^m$ as
\begin{equation}
    p(k^m) \approx \frac{1}{|{\mathcal{D}}|} \sum_{x\in {\mathcal{D}}} p(k^m|x),
\end{equation}
where ${\mathcal{D}}$ denotes the training dataset. Then, by definition, the entropy of the codeword index $k^m$ can be estimated as
\begin{equation}
    H(K^m) = - \sum_{k^m=1}^K p(k^m) \log\left(p(k^m)\right).
\end{equation}
Similarly, the conditional entropy of codeword $k^m$ given data $x$ can be estimated as
\begin{equation}
    H(K^m|X) \! = \! - \frac{1}{|{\mathcal{D}}|} p(k^m|x) \log \!\left(p(k^m|x)\right).\!
\end{equation}
Now, the mutual information between codeword index $k^m$ and data $x$ can be easily obtained by definition as
$I(X, K^m) = H(K^m) - H(K^m|X)$. In practice, we find that it is better not to directly maximize the MI $I(X, K^m)$, but to maximize its variant form
\begin{equation}
    I(X, K^m) = H(K^m) - \alpha H(K^m|X),
    \label{fml:mi}
\end{equation}
where $\alpha$ is a non-negative hyper-parameter controlling the trade-off between two entropy terms. Intuitively, maximizing the MI can be understood as encouraging only one codeword is assigned a high probability for a document $x$, while all codewords are used evenly overall. Given the mutual information, the final training objective becomes
\begin{equation}
    {\mathcal{L}} = \overline {\mathcal{L}}_{cl} - \lambda \sum_{m=1}^M I(X, K^m),
    \label{fml:overall}
\end{equation}
where $\lambda$ is a hyper-parameter controlling the relative importance of the MI term. Since this method employs MI to improve the quality of codewords, we name the model as \textbf{MICPQ}, i.e., \textbf{M}utual-\textbf{I}nformation-Improved \textbf{C}ontrastive \textbf{P}roduct \textbf{Q}uantization.

\section{Related Work}
As the main solution to efficient document retrieval, unsupervised semantic hashing has been studied for years. Many existing unsupervised hashing methods are established on the generative models encouraging the binary codes to reconstruct the original document. For example, VDSH \citep{ChaidaroonF17} proposes a two-stage scheme, in which it first learns the continuous representations under the VAE \citep{KingmaW13} framework, and then cast them into the binary codes. To tackle the two-stage training issue, NASH \citep{HenaoCSSWWC18} presents an end-to-end generative hashing framework where the binary codes are treated as Bernoulli latent variables, and introduces the Straight-Through \citep{BengioLC13} estimator to estimate the gradient w.r.t. the discrete variables. \citet{DongSSC19} employs the mixture priors to empower the binary code with stronger expressiveness, therefore resulting in better performance. Further, CorrSH \citep{ZhengSSC20} employs the distribution of the Boltzmann machine to introduce correlations among the bits of binary codes. \citet{YeMY20} proposes the auxiliary implicit topic vectors to address the issue of information loss in the few-bit scenario. Also, a handful of recent works focus on how to inject the neighborhood information of the graph under the VAE framework. \citet{ChaidaroonEF18, Hansen0SAL20a} seeks to learn the optimal binary codes that can reconstruct neighbors of original documents. A ranking loss is introduced in \citep{Hansen0SAL19} to accurately characterize the correlation between documents. \citet{OuSYLWZCZ20} first proposes to integrate the semantics and neighborhood information with a graph-driven generative model.

Beyond generative hashing methods, studies on hashing via the mutual information (MI) principle emerges recently. AMMI \citep{StratosW20} learns a high-quality binary code by maximizing the MI between documents and binary codes. 
DHIM \citep{OuSYZZL21} first compresses the  BERT embeddings into binary codes by maximizing the MI between global codes and local codes from documents.

Another efficient retrieval mechanism is product quantization. The Product quantization (PQ) \citep{JegouDS11} and its improved variants such as Optimized PQ \citep{GeHK013} and Locally Optimized PQ \citep{KalantidisA14} are proposed to retrain richer distance information than hashing methods while conducting the retrieval efficiently. These shallow unsupervised PQ methods are often based on the well-trained representation and learn the quantization module with heuristic algorithms \citep{xiao2021matching}, which often can not achieve satisfactory performance. In this paper, we propose an unsupervised MI-improved end-to-end unsupervised product quantization model MICPQ. We notice that the proposed MICPQ is somewhat similar to recent works w.r.t PQ \citep{JangC21,abs-2109-05205}  in the computer vision community. However, \cite{JangC21} focuses on analyzing the performance difference between different forms of contrastive losses, while we focus on how to design an end-to-end model to jointly refine and quantize the BERT embedding via contrastive product quantization from a probabilistic perspective. \cite{abs-2109-05205} concentrates on how to improve the  codeword diversity to prevent model degeneration by regularizer design; whereas there is no "model degeneration" phenomenon observed in our model, and we use the mutual information maximization based method to increase the representativeness of codewords to further improve the retrieval performance.

\section{Experiments}
\label{experiment}
\subsection{Datasets, Evaluation and Baselines}
\paragraph{Datasets} 
The proposed MICPQ model is evaluated on three benchmark datasets, including NYT \citep{TaoZCJHK018}, AGNews \citep{ZhangZL15} and DBpedia \citep{LehmannIJJKMHMK15}. Details of the three datasets can be found in Appendix \ref{app:datasets}.

\paragraph{Evaluation Metrics}
For every query document from the testing dataset, we retrieve its top-100 most similar documents from the training set with the Asymmetric distance computation \citep{JegouDS11} which is formulized in Appendix \ref{app:retrieval}. Then the retrieval precision is calculated as the ratio of the relevant documents. Note that a retrieved document is considered relevant to the query if they both share the same label. Finally, the retrieval precision averaged over all test documents is reported.

\paragraph{Baselines}
We consider the following unsupervised deep semantic hashing methods for comparision: VDSH \citep{ChaidaroonF17}, NASH \citep{HenaoCSSWWC18}, BMSH \cite{DongSSC19}, CorrSH \citep{ZhengSSC20}, WISH \citep{YeMY20}, AMMI \citep{StratosW20}, DHIM \citep{OuSYZZL21}. For the reported performances of baselines, they are quoted from DHIM \citep{OuSYZZL21}.
Apart from existing baselines, we also implement the following two baselines for a more thorough comparison.

\textbf{AEPQ}. We are interested in optimizing our product quantizer with the reconstructing objective. Specifically, the quantized semantics-preserving representation vector $h(x)$ is expected to reconstruct the original input features (i.e., BERT embeddings) with a newly added decoder, which is similar with previous generative hashing methods. Same as MICPQ, the performance of AEPQ is evaluated with the Asymmetric distance computation.
We name this baseline as AEPQ, i.e., \textbf{A}uto-\textbf{E}ncoder-based \textbf{P}roduct \textbf{Q}uantization.

\textbf{CSH}.  Same as the encoder setting in NASH \citep{HenaoCSSWWC18}, we assume that the binary codes are generated by sampling from the multivariate Bernoulli distribution and propose to minimize the expected contrastive loss w.r.t. the discrete binary codes directly. The straight-through (ST) gradient estimator \citep{BengioLC13} is utilized for training in an end-to-end manner. The data augmentation strategy is the same as that of our MICPQ. We name the proposed baseline as CSH, i.e., \textbf{C}ontrastive \textbf{S}emantic \textbf{H}ashing.

\begin{table*}[!t]
    \caption{The precision(\%) comparision with different state-of-the-art unsupervised efficient retrieval methods. Among them, bold numbers represent best performance, and underlined numbers represent second best performance.}
    \centering
    \resizebox{\textwidth}{!}{
    \begin{tabular}{c|cccc|cccc|cccc}
    \hline 
    \multirow{2}*{\textbf{Method}} & \multicolumn{4}{c}{\textbf{NYT}}  &\multicolumn{4}{|c}{\textbf{AGNews}} & \multicolumn{4}{|c}{\textbf{DBpedia}} \\
    
    ~      & 16bits & 32bits & 64bits & 128bits & 16bits & 32bits & 64bits & 128bits & 16bits & 32bits & 64bits & 128bits  \\
    \hline 
    \hline 
    \multicolumn{13}{c}{Using TFIDF features} \\ 
    \hline 
    VDSH    &68.77  &68.77  &75.01  &78.49  &67.32  &67.42  &72.70  &73.86  &67.79  &72.64  &78.84  &84.91     \\
    NASH    &74.87  &75.52  &75.08  &73.01  &65.74  &69.34  &72.72  &74.33  &78.02  &79.84  &79.79  &76.76      \\
    WISH    &70.15  &70.03  &64.48  &68.94  &74.53  &74.79  &75.05  &72.70  &82.82  &82.76  &82.10  &78.22      \\
    BMSH    &74.02  &76.38  &76.88  &77.63  &74.09  &76.03  &76.09  &73.56  &83.17  &86.24  &87.05  &83.86      \\
    CorrSH  &75.43  &77.61  &77.24  &78.39  &76.20  &76.45  &76.61  &77.67  &82.01  &81.78  &80.94  &85.77      \\
    AMMI    &71.06  &76.48  &77.37  &78.03  &76.47  &76.61  &77.32  &78.23  &84.51  &89.53  &90.78  &91.03      \\
    \hline 
    \hline 
    \multicolumn{13}{c}{Using BERT [CLS] embeddings} \\ 
    \hline 
    VDSH    &53.38  &58.18  &62.44  &64.64  &62.97  &66.35  &69.57  &70.27  &69.59  &75.21  &79.54  &80.62      \\
    NASH    &55.87  &58.25  &60.98  &64.27  &66.32  &68.44  &70.40  &72.07  &65.87  &74.54  &77.96  &81.43      \\
    WISH    &58.83  &64.75  &65.47  &70.34  &65.35  &66.19  &69.39  &72.03  &65.65  &72.91  &76.66  &82.29      \\
    BMSH    &59.35  &63.26  &65.87  &69.71  &66.77  &69.61  &71.99  &73.16  &66.42  &79.13  &82.01  &84.57      \\
    CorrSH  &62.03  &65.48  &68.38  &72.28  &67.06  &68.51  &70.86  &73.17  &65.28  &74.63  &78.65  &83.61      \\
    AMMI    &60.47  &65.10  &69.67  &74.47  &65.50  &68.26  &71.85  &74.36  &80.25  &82.67  &89.26  &86.74      \\
    DHIM    &\underline{79.69}  &80.55  &79.77  &79.09  &78.23  &79.17  &78.88  &79.86  &94.26  &94.80  &93.02  &88.21      \\
    \textbf{AEPQ}    &67.98  &68.35  &70.38  &71.48  &68.81  &70.49  &72.82  &73.84  &80.24  &84.54  &86.78  &89.70  \\
    \textbf{CSH}    &79.63  &\underline{81.05}  &\underline{81.08}  &\underline{81.28}  &\underline{79.14}  &\underline{80.25}  &\underline{81.14}  &\underline{81.78}  &\underline{95.19}  &\underline{95.84}  &\underline{95.79}  &\underline{96.01}      \\
    \textbf{MICPQ}     &\textbf{83.15}  &\textbf{84.02}  &\textbf{84.24}  &\textbf{85.08}  &\textbf{80.36}      &\textbf{81.84}      &\textbf{82.93}      &\textbf{83.29}      &\textbf{96.51}  &\textbf{97.00}  &\textbf{97.03}  &\textbf{97.21}      \\

    \hline 
    
    \end{tabular}
    }
    \vspace{-2mm}
    \label{tab:main}
\end{table*}

\paragraph{Training Details}
In our MICPQ, to output the refined vectors that will be fed into the product quantizer with the desired dimension, the encoder network is constituted by a pre-trained BERT Module followed by an one-layer ReLU feedforward neural network on top of the [CLS] representation whose dimension is 768. For performances under the average pooling setting of BERT, please refer to Appendix \ref{app:pooling}. During the training, following the setting in DHIM \citep{OuSYZZL21}, we fix the parameters of BERT, while only training the newly added parameters. We implement the proposed model with PyTorch and employ the Adam optimizer\citep{KingmaB14} for optimization.

In terms of hyper-parameters relevant to the product quantization, the dimension of  codeword $\frac{D}{M}$ in the small codebook $C^m$ is fixed to $24$ and the number of codewords $K$ in each codebook $C^m$  is fixed to $16$. By setting the number of small codebooks $M$ as $\{4, 8, 16 ,32 \}$, we can see that the final codeword in the codebook $C$ can be represented by $\{16, 32, 64, 128 \}$ bits according to $B = M log_2 K$. Thus, when compared with semantic hashing, they are compared under the same number of bits.

For other hyper-parameters, the learning rate is set as $0.001$; the dropout rate $p_{drop}$ to generate positive pairs for contrastive learning is set as 0.3; the Gumbel-Softmax temperature $\tau$ is set as $10$ for 16-bit binary codes and $5$ for longer codes; the temperature $\tau_{cl}$ in  contrastive learning is set as $0.3$; the trade-off coefficient $\alpha$ in \eqref{fml:mi} is set as $0.1$; the coefficient $\lambda$ in \eqref{fml:overall} is chosen from $\{0.1,0.2, 0.3\}$ according to the performance on the validation set. 


\subsection{Results and Analyses}

\subsubsection{Overall Performance}

Table \ref{tab:main} presents performances of our proposed model and existing baselines on three public datasets with code lengths varying from 16 to 128. It can be seen that the proposed simple baseline CSH achieves promising performances across all three datasets and nearly all code length settings when compared to previous state-of-the-art methods, demonstrating the superiority of using contrastive learning to promote semantic information. Further, our proposed MICPQ  outperforms the previous methods by a more substantial margin. Specifically,  improvements of $4.32\%$, $3.07\%$ and $4.37\%$ averaged over all code lengths are observed on NYT, AGNews and DBpedia datasets, respectively,  when compared with the current state-of-the-art DHIM. Moreover, the performance of AEPQ lags behind our proposed MICPQ remarkably, which illustrates the limitation of the reconstructing objective. It is also observed that the retrieval performance of our proposed MICPQ consistently improves as the code length increases. Although this is consistent with our intuition that a longer code can preserve more semantic information, it does not always hold in some previous models (e.g., NASH, BMSH, DHIM).

\subsubsection{Abalation Study}
To understand the influence of different components in the MICPQ, we further evaluate the retrieval performance of two variants of MICPQ. \emph{(i)} $\text{MICPQ}_{\text{cl}}$: it removes the mutual-information term $I(X,K^m)$ in each codebook and only optimizes the quantized contrastive loss to learn the semantics-preserving quantized representation; 
\emph{(ii)} $\text{MICPQ}_{\text{softmax}}$: it does not inject the Gumbel noise, but only utilizes the sole softmax operation to produce the deterministic codeword index. As seen from Table \ref{tab:component}, when compared to $\text{MICPQ}_{\text{cl}}$, our MICPQ improves the retrieval performance averaged over all code lengths by $1.51\%$ and $0.94\%$ on NYT and AGNews respectively, demonstrating the effectiveness of our mutual-information term inside each codebook. Also, by comparing MICPQ to $\text{MICPQ}_{\text{softmax}}$, consistent improvements can be observed on both datasets, which demonstrates the superiority of the proposed probabilistic product quantization.

\begin{table}[!t]
    \caption{The precision (\%) comparision with variants of MICPQ.}
    \vspace{-1mm}
    \centering
    \resizebox{\linewidth}{!}{
    \begin{tabular}{c|c|cccc}
    \hline
    \multicolumn{2}{c|}{\textbf{Ablation Study}}    &16bits   &32bits   &64bits   &128bits \\
    \hline
    \hline 
    \multirow{3}*{NYT}  & $\text{MICPQ}_{\text{cl}}$     &81.77  &82.28  &82.96  &83.63 \\ 
    ~                   & $\text{MICPQ}_{\text{softmax}}$     &82.28  &83.46  &83.64  &84.17 \\ 
    ~                   & MICPQ       &\textbf{83.15}  &\textbf{84.02}  &\textbf{84.24}  &\textbf{85.08} \\ 
    \hline 
    \hline 
    \multirow{3}*{AGNews}   & $\text{MICPQ}_{\text{cl}}$    &79.68  &80.97  &81.79  &82.23 \\ 
        ~                   & $\text{MICPQ}_{\text{softmax}}$    &79.60  &81.61  &82.58  &82.69 \\ 
        ~                   & MICPQ       &\textbf{80.36}  &\textbf{81.84}  &\textbf{82.93}  &\textbf{83.29} \\ 
    \hline
    \end{tabular}
    }
    \label{tab:component}
\end{table}

\subsection{Link to Semantic Hashing: A Special Form of MICPQ}
\begin{figure}[!t]
    \centering
    \includegraphics[width = 0.8\columnwidth]{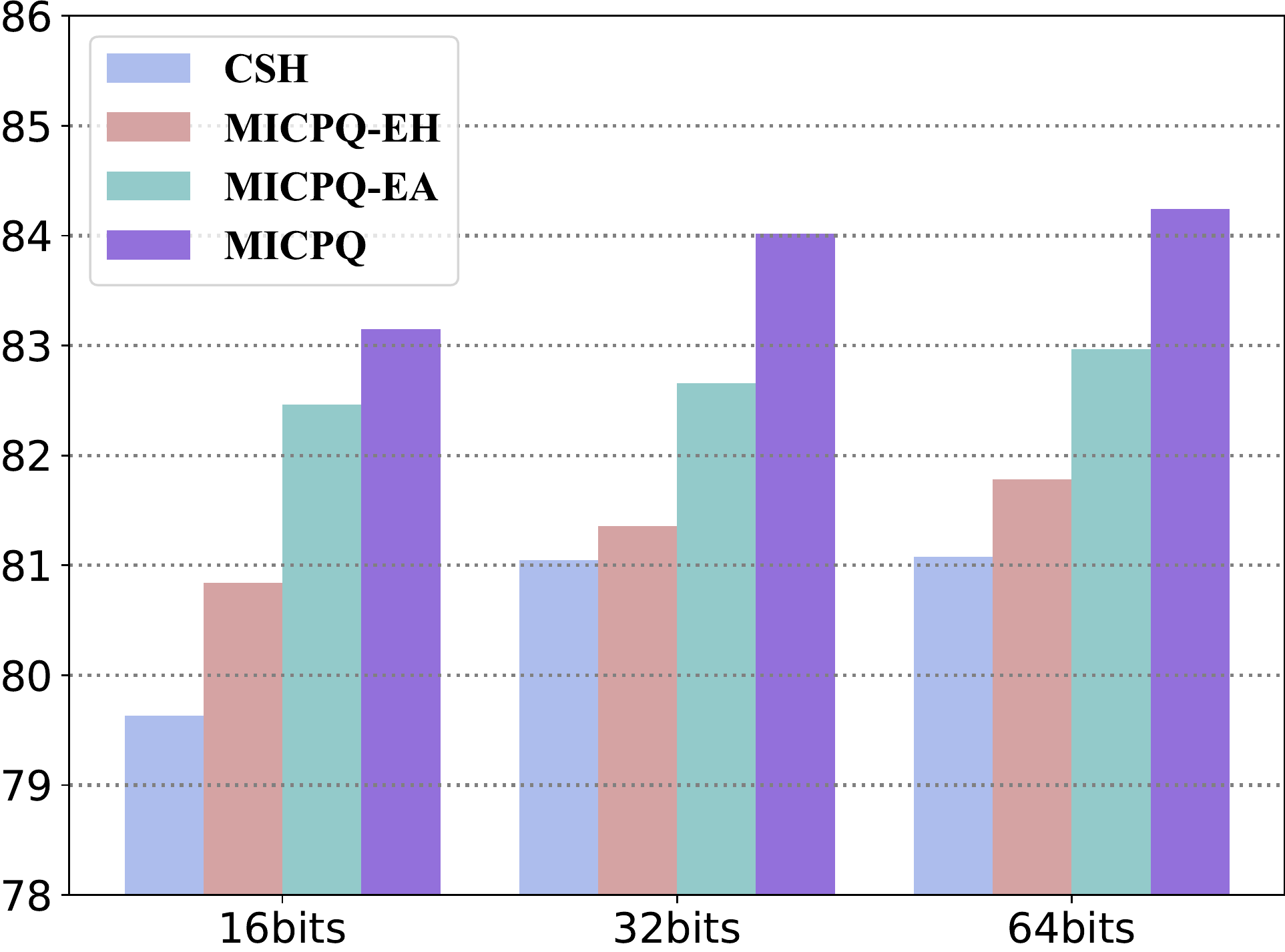}
    \caption{The precision(\%) on NYT  w.r.t. MICPQ under the extreme setting.}
    \label{fig:bridge}
    \vspace{-2mm}
\end{figure}

Through an extreme setting, the MICPQ model can be reconsidered in the semantic hashing framework and be evaluated using the Hamming distance rather than the Asymmetric distance. Specifically, we push the number of codebooks $M$ equal to its maximal limit (i.e., the code length $B$), and the number of codewords $K$ inside each codebook is forced as 2 to satisfy the equation $B = M \log_2K$. This way, the state of each bit in the $B$-bit binary code will be decided by a sub-space with 2 codewords. Under this extreme setting, we can either  evaluate MICPQ with the Hamming distance evaluation or the Asymmetric distance. We name both models as MICPQ-EH (i.e., \textbf{E}xtreme-\textbf{H}amming) and MICPQ-EA (i.e., \textbf{E}xtreme-\textbf{A}symetric). As shown in Figure \ref{fig:bridge}, the MICPQ-EA consistently outperforms the MICPQ-EH thanks to the superiority of  Asymmetric Distance computation. Also, MICPQ-EH can be seen as the semantic hashing model since it's evaluated with Hamming distance, and the better retrieval performance of MICPQ-EH when compared to CSH informs us that we can take the special form of MICPQ as one of the excellent semantic hashing methods. To sum up, the proposed MICPQ is more flexible and powerful than the existing hashing baselines in terms of efficient document retrieval.

\begin{table}[!t]
    \caption{Average and maximal accuracy(\%) over $M$ codebooks on NYT.}
    \centering
    \resizebox{\linewidth}{!}{
    \begin{tabular}{c|ccc|ccc}
        \hline
         ~ & \multicolumn{3}{c|}{Avg Acc}  & \multicolumn{3}{|c}{Max Acc} \\
         ~ & NYT    & AGNews    & DBpedia   & NYT    & AGNews    & DBpedia \\     
         \hline
         KMeans &41.73 &\textbf{61.10}   &\textbf{67.99} &43.74  &67.45  &80.49  \\
         MICPQ   &\textbf{48.74}  &58.06  &64.12  &\textbf{57.51}  &\textbf{69.65}  &\textbf{80.88} \\ 
         \hline
    \end{tabular}
    }
    \label{tab:acc}
\end{table}

\subsection{Evaluating the Quality of Codewords from the Clustering Perspective}
To examine how well the codewords can represent the corresponding refined vectors $ \tilde z^m(x)$, we set the number of codewords $K$ in each codebook as the number of the ground-truth classes on datasets (i.e., $K = 26, 14 $ and  $4$ on NYT, DBpedia and AGNews respectively), so that we can compute the unsupervised clustering accuracy with the help of the Hungarian algorithm \citep{kuhn1955hungarian} in each codebook. The number of codebooks $M$ is set as $8$ on all datasets. For comparison, we also run K-Means on each $ \tilde z^m(x)$ separately. As shown in Table \ref{tab:acc}, the codewords in our MICPQ are on par with the ones learned by K-Means.  Particularly, our MICPQ significantly outperforms  K-means on NYT that has the largest number of classes (i.e., 26), demonstrating the ability of our MICPQ to learn high-quality codewords, especially for datasets with diverse categories.

\begin{figure}[!t]
    \centering
	\subfigure{
		\begin{minipage}[b]{0.45\columnwidth}
  	 	\includegraphics[width = 1\columnwidth]{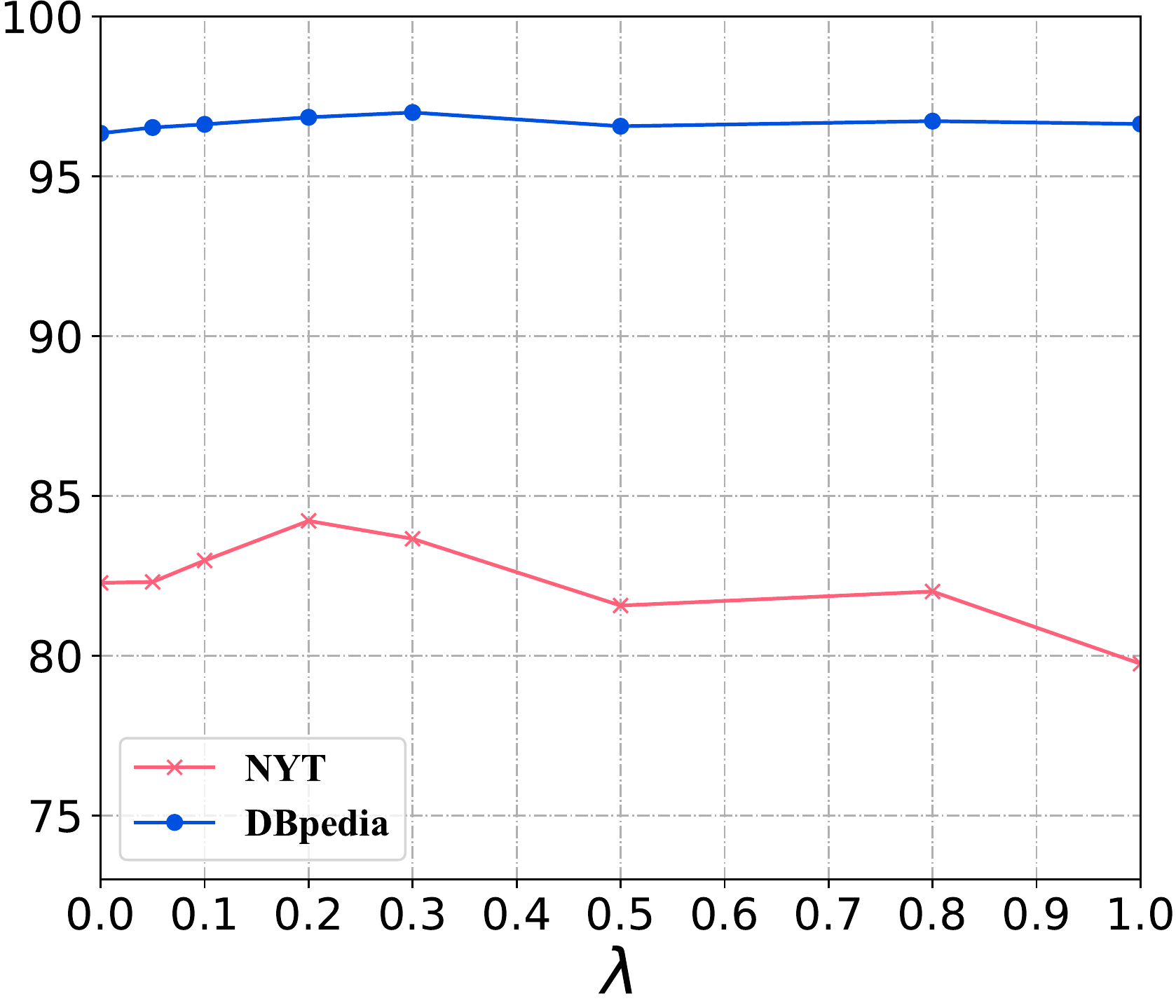}
		\end{minipage}
	}
	\subfigure{
		\begin{minipage}[b]{0.45\columnwidth}
  	 	\includegraphics[width = 1\columnwidth]{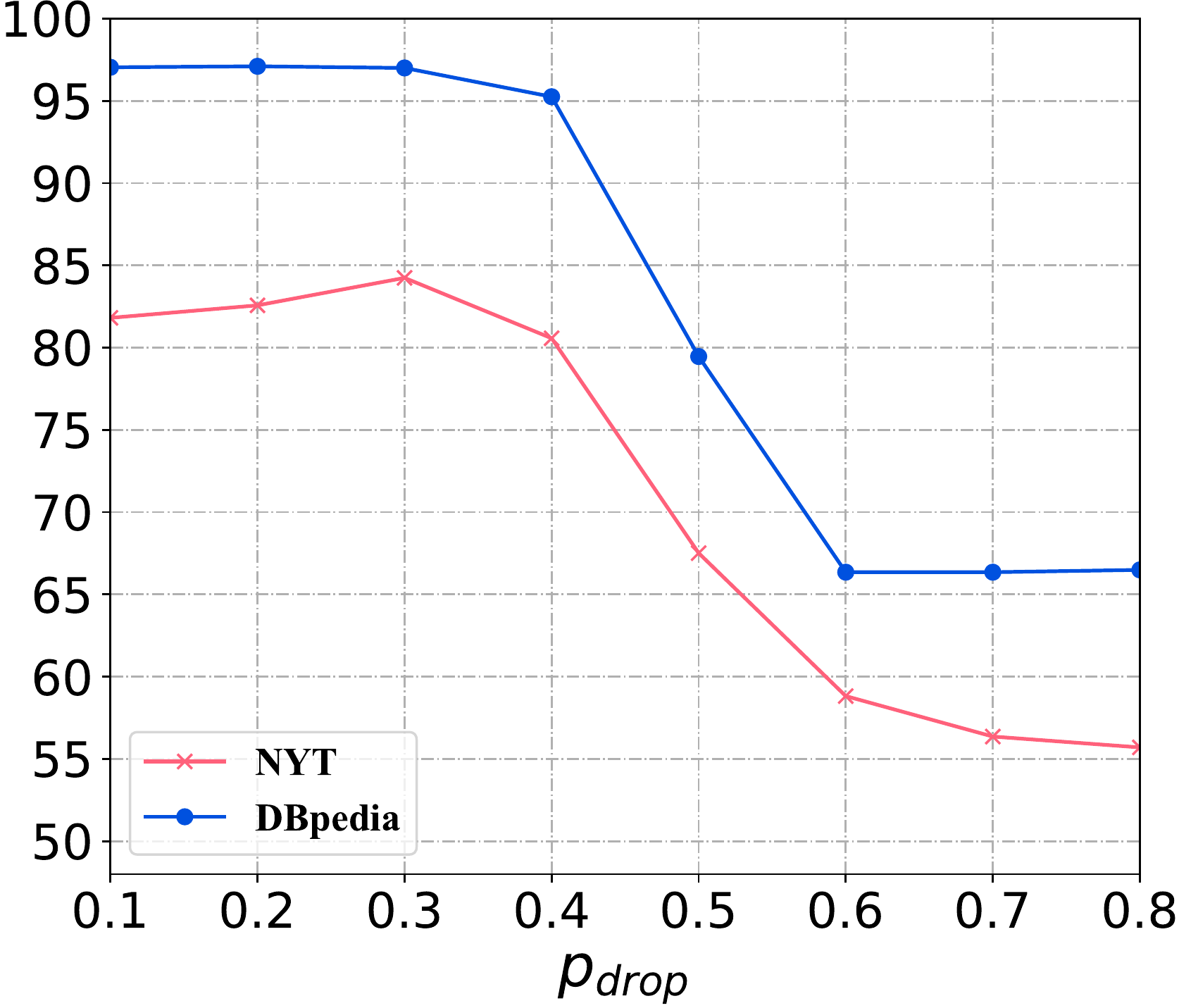}
		\end{minipage}
	}
	\subfigure{
		\begin{minipage}[b]{0.45\columnwidth}
  	 	\includegraphics[width = 1\columnwidth]{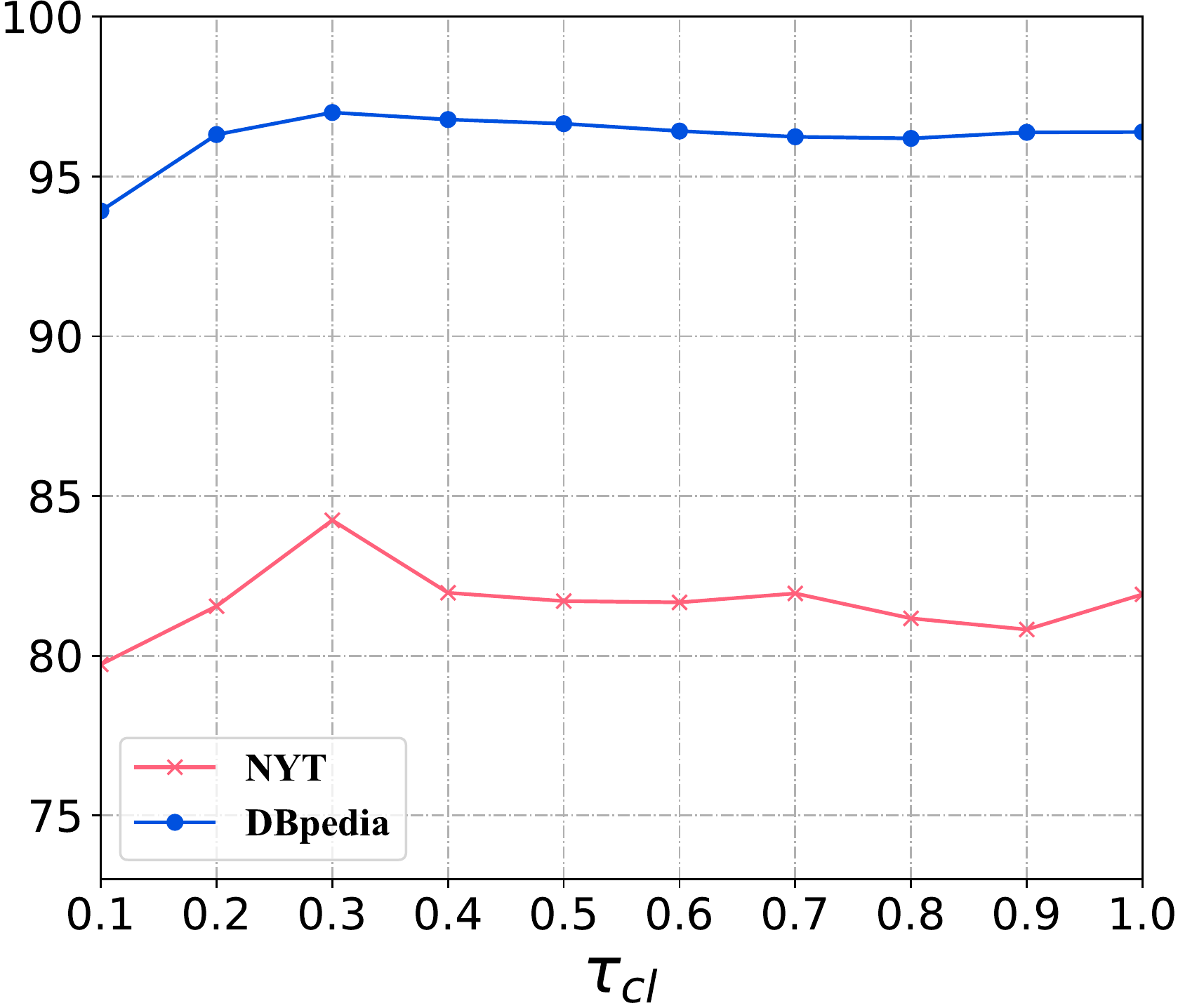}
		\end{minipage}
	}
	\subfigure{
		\begin{minipage}[b]{0.45\columnwidth}
  	 	\includegraphics[width = 1\columnwidth]{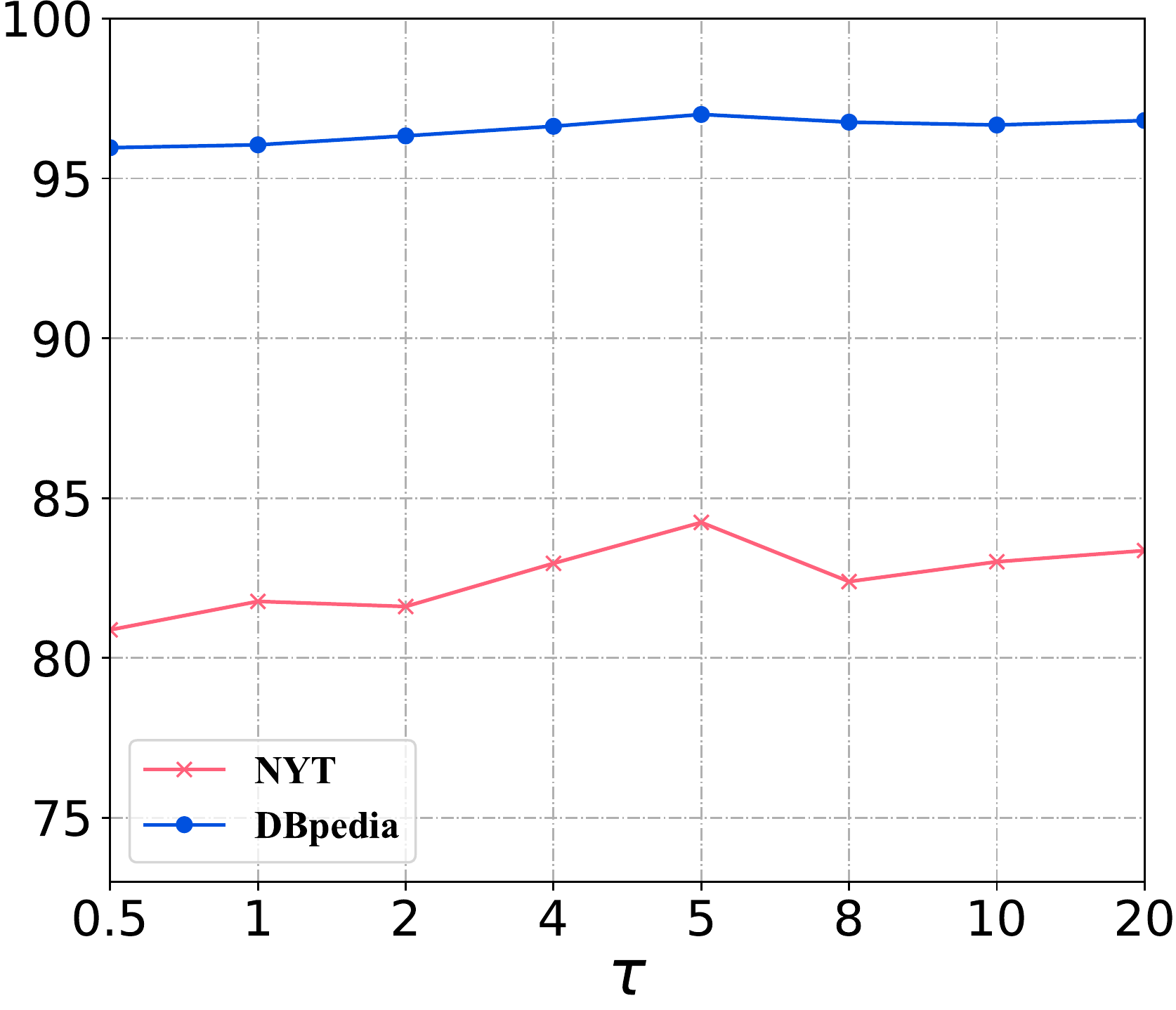}
		\end{minipage}
	}
	\caption{Hyper-parameter analyses with the precision(\%) of 32-bits setting on NYT and DBpedia.}
	\label{fig: hypersensitivity}
\end{figure}

\subsection{Sensitive Analyses of Hyper-Parameters}
We investigate the influence of 4 key hyper-parameters: the coefficient $\lambda$, the dropout rate $p_{drop}$, the temperature $\tau_{cl}$ in contrastive learning, and the Gumbel-Softmax temperature $\tau$. 
As shown in Figure \ref{fig: hypersensitivity}, compared with the case of $\lambda = 0$, obvious performance gains can be observed by introducing the mutual-information loss inside each codebook when the $\lambda$ is set as a relatively small value (e.g., 0.2 or 0.3). The value of $p_{drop}$ controls the strength of data augmentation. We see that as $p_{drop}$ exceeds some thresholds (e.g., 0.4), the performance will decrease sharply, and eventually the model will collapse on both datasets. Also, it is shown that as $\tau_{cl}$ grows up, the precision first increases and reaches the peak when $\tau_{cl}$ is around $0.3$ on both datasets. As for the Gumbel-Softmax temperature $\tau$, we suggest setting it to a larger value, saying [5,15].

\section{Conclusion}

In this paper, we proposed a novel unsupervised product quantization model, namely MICPQ. In MICPQ, we managed to develop an end-to-end probabilistic contrastive product quantization model to jointly refine the original BERT embeddings and quantize the refined embeddings into codewords, with a probabilistic contrastive loss designed to preserve the semantic information in the quantized representations. Moreover, to improve the representativeness of codewords for keeping the cluster structure of documents soundly, we proposed to maximize the mutual information between data and the codeword assignment. Extensive experiments showed that our model significantly outperformed existing unsupervised hashing methods.

\section{Limitations}
In our work, we do not analyze the individuality of codebooks and the difference between codebooks. However, each codebook should show different aspects of characteristics for preserving rich semantics. Therefore, it may help if the difference between codebooks is more analyzed and improved this way. Also, in the future we may try to apply this method to the tasks of passage retrieval, on which some large-scale datasets for evaluation are available.

\section*{Acknowledgement}
This work is supported by the National Natural Science Foundation of China (No. 62276280, U1811264), Key R\&D Program of Guangdong Province (No. 2018B010107005), National Natural Science Foundation of Guangdong Province (No. 2021A1515012299), Science and Technology Program of Guangzhou (No. 202102021205).  This work is also supported by CAAI-Huawei MindSpore Open Fund.

\bibliography{anthology,custom}
\bibliographystyle{acl_natbib}

\clearpage

\appendix
\section{Details about Datasets}
\label{app:datasets}

\begin{table}[htbp]
    \caption{Statistics of Three Benchmark Datasets. }
    \centering
\resizebox{\linewidth}{!}{
    \begin{tabular}{cccccc}
    \hline 
         \textbf{Dataset} & \textbf{Classes} & \textbf{AvgLen} & \textbf{Train} & \textbf{Val} & \textbf{Test}  \\
    \hline
         NYT     & $26$ & $648$ & $9,221$ & $1,154$ & $1,152$   \\
         AGNews  & $4$  & $32$  & $114,839$ & $6,381$ & $6,380$ \\ 
         DBpedia & $14$ & $47$  & $50,000$ & $5,000$ & $5,000$ \\
    \hline
    \end{tabular}
    }
    \label{tab:statistics}
\end{table}

Three datasets are used to evaluate the performance of the proposed model.  \emph{1) NYT} \citep{TaoZCJHK018} is a dataset which contains news articles published by The New York Times; \emph{3) AGNews}  \citep{ZhangZL15} is a news collection gathered from academic news search engines; \emph{3) DBpedia} \citep{LehmannIJJKMHMK15} is a dataset which contains the abstracts of articles from Wikipeida. We simply apply the string cleaning operation same as in \citep{OuSYZZL21}. The statistics of the three datasets are shown in Table \ref{tab:statistics}.

\section{Retrieval}
\label{app:retrieval}
For testing, we first encode all the documents in the search set (i.e., the training set in our setting). For each document $x_i$ in the search set, we encode it using the hard quantization operation to obtain the codeword index $\{k^1_i, k^2_i, \cdots, k^M_i \}$. Then all these codeword indices are concatenated and stored in the form of the $M \log_2K$-bits binary code.

In terms of the retrieval stage, when given a query document $x_q$, we first extract its refined embedding $\tilde z(x_q)$ through the encoder network and then slice it into $M$ equal-length segments as $\tilde z(x_q) = [\tilde z^1(x_q), \tilde z^1(x_q), \cdots, \tilde z^M(x_q) ]$. Then we compute the Asymmetric distance (AD) \citep{JegouDS11} between the query $x_q$ and the documents $x_i$ with the squared Euclidean distance metric as:
\begin{equation}
    \!AD(x_q, x_i)\! = \!\sum^M_{m = 1} \! ||\tilde z^m(x_q) - C^m \cdot h^m(x_i)  ||^2_2,\!
\end{equation}
where $h^m(x_i) = C^m \cdot one\_hot(k_i^m)$ represents the quantized representation (i.e., one of codewords) of $x_i$ in the $m$-th codebook. To compute that, we can first pre-compute a query-specific distance look-up table of size $M \times K$ that stores the distance between the segment $\tilde z^m(x_q)$ and all codewords in each codebook. With the pre-computed look-up table, $AD(x_q,x_i)$ can be efficiently computed by summing up the chosen values from the look-up table. It is only slightly more costly when compared with the efficiency of the Hamming distance.

\section{Pooling methods}

\begin{table}[htbp]
    \caption{The performance comparision between the [CLS] representations and the average embeddings of BERT.}
    \centering
    \resizebox{\linewidth}{!}{
    \begin{tabular}{c|c|cccc}
    \hline
    \multicolumn{2}{c|}{\textbf{CLS vs Avg}}       &16bits   &32bits   &64bits   &128bits \\
    \hline
    \hline 
    \multirow{4}*{NYT}  & $\text{CSH}_{\text{Avg}}$      &71.89  &71.99  &77.38  &78.26 \\
    ~                   & $\text{CSH}_{\text{CLS}}$      &79.63  &81.05  &81.08  &81.28 \\
    ~                   & $\text{MICPQ}_{\text{Avg}}$     &79.56  &79.99  &78.49  &80.33 \\ 
    ~                   & $\text{MICPQ}_{\text{CLS}}$         &\textbf{83.15}  &\textbf{84.02}  &\textbf{84.24}  &\textbf{85.08} \\ 
    \hline 
    \hline 
    \multirow{4}*{AGNews}   & $\text{CSH}_{\text{Avg}}$    &81.38  &81.22  &81.83  &81.85 \\ 
    ~                   & $\text{CSH}_{\text{CLS}}$      &79.14  &80.25  &81.14  &81.78 \\
    ~                    & $\text{MICPQ}_{\text{Avg}}$    &\textbf{81.60}  &\textbf{83.00}  &\textbf{83.85}  &\textbf{84.04} \\  
    ~                   & $\text{MICPQ}_{\text{CLS}}$       &80.36  &81.84  &82.93  &83.29 \\ 
    \hline
    \hline
    \multirow{4}*{DBpedia} & $\text{CSH}_{\text{Avg}}$ &93.39 &93.98  &94.56  &94.66  \\
    ~                   & $\text{CSH}_{\text{CLS}}$      &95.19  &95.84  &95.79  &96.01  \\
    ~                    & $\text{MICPQ}_{\text{Avg}}$ &94.47  &95.87  &95.76  &95.71  \\
    ~                       & $\text{MICPQ}_{\text{CLS}}$ &\textbf{96.51}  &\textbf{97.00}  &\textbf{97.03}  &\textbf{97.21}  \\
    \hline
    \end{tabular}
    \label{tab:pooling}
    }
    
\end{table}

\label{app:pooling}
We are interested in if taking the average embeddings of the last layers in BERT as inputs will lead to a better performance than [CLS] or not. The experiments are conducted in our proposed model MICPQ and the simple baseline CSH. Table \ref{tab:pooling} shows that these two settings achieve better performance in different datasets respectively. For simplicity, we consistently use the [CLS] representation in our experiments.


\end{document}